\documentclass[12pt,aip,pof]{revtex4-1}
\usepackage[latin9]{inputenc}
\usepackage{amsmath}
\usepackage{graphicx}

\makeatletter

\usepackage{epsfig,dcolumn,bm}

\def\vec#1{\boldsymbol{\mathrm{#1}}}

\makeatother

\begin{document}
\title{Ponomarenko dynamo sustained by a free swirling jet 
}

\author{I. Grants}
\author{J. Priede}
\altaffiliation[Also at ]{Fluid and Complex Systems Research Centre, Coventry University, Coventry, CV1 5FB, UK}
\affiliation{Faculty of Science and Technology, University of Latvia, Riga, LV-1004, Latvia}

\begin{abstract}
We present numerical results on dynamo action in a flow driven by
an azimuthal body force localized near the end of an elongated cylindrical
container. The analysis focuses on the central region of the cylinder,
where axial variations in the flow are relatively weak, allowing the
magnetic field to be represented as a helically traveling wave. Four
magnetic impeller configurations and multiple forcing intensities
are examined. In all cases, the velocity profiles in the central region
display a similar $\propto r^{-2}$ dependence across a wide range
of Reynolds numbers and forcing region widths. The magnetic field is found
to start growing under conditions similar to those of the Riga dynamo.
However, the growing modes exhibit a substantial nonzero group velocity,
indicating that the associated instability is convective: the flow
can amplify an externally applied magnetic field but cannot sustain
it autonomously. We outline several approaches for overcoming this
limitation in order to realize a working laboratory dynamo based on
an internally unconstrained swirling jet-type flow.
\end{abstract}
\maketitle

\section{Introduction}

Screw-like motion of an electrically conducting fluid is arguably the simplest flow capable of sustaining a magnetic field through dynamo action.\citep{Lortz1968} The most prominent example is the Ponomarenko dynamo,\citep{Ponomarenko1973} which can generate a magnetic field at relatively low flow velocities attainable in laboratory conditions.\citep{Gailitis1976} This was first demonstrated in the celebrated Riga dynamo experiment \citep{Gailitis2000,Gailitis2001}, where the flow of liquid sodium was kinematically constrained and guided to mimic the solid-body helical motion of the original Ponomarenko dynamo. An even more constrained configuration was used in the Karlsruhe dynamo experiment, which, however, relied on a different dynamo model, the Roberts-Busse dynamo.\citep{Stieglitz2001} In both cases, the flow had severely restricted freedom to respond to the growing electromagnetic force once the magnetic field reached significant strength. This nonlinear interaction between the flow and the magnetic field -- responsible not only for magnetic field saturation but also for potentially complex temporal behavior -- is arguably the most scientifically challenging aspect of fluid dynamos.

The aim of the present study is to numerically explore the feasibility of a laboratory screw-type dynamo driven by an impeller in a cylindrical vessel, whose only constraints are the external walls. Previous analysis based on Wentzel-Kramers-Brillouin (WKB) approximation suggests that a smooth swirling-jet flow, which is often referred to as the smooth Ponomarenko dynamo,\citep{Gilbert2003} may be able to generate a magnetic field at significantly lower velocities than the solid-body Ponomarenko dynamo.\citep{Ruzmaikin1988} However, the accuracy of the underlying asymptotic solution is uncertain, and the analysis relies on highly idealized velocity profiles.

In the present study, we numerically solve the one-dimensional induction
equation for velocity profiles that approximate a concentrated vortex
driven by a small-diameter impeller in a finite cylinder. A striking
and well-known example of such a flow is produced by a laboratory
magnetic bar stirrer.\citep{Halasz2007} A similar vortex flow can
also be generated by azimuthal electromagnetic body forces arising
from a rotating permanent magnet placed coaxially near the cylinder's
end wall.\citep{Grants2021} We use direct numerical simulations
to compute several representative realizations of such flows, which
then serve as input for the induction equation within the smooth Ponomarenko
dynamo framework.

The flow described above may enable a technically simple implementation
of a liquid-metal laboratory dynamo using a commercially manufactured
large sodium storage tank, which can contain up to 22 m$^{3}$ of
liquid sodium. In the simplest realization, the flow could be driven
in an unmodified tank by a rotating permanent magnet. Alternatively,
the flow could be generated by a mechanically driven impeller, actuated
either through a sealed shaft or by magnetic coupling.

The considered configuration shares strong similarities with the Riga
dynamo.\citep{Gailitis2000,Gailitis2001} The primary distinction
is the absence of internal walls, which results in smooth radial velocity
profiles. Eliminating these walls simplifies the experiment by removing
the requirement for reliable electrical contact across them. It also
relaxes kinematic constraints on how the flow can respond to the generated
magnetic field. Consequently, the proposed configuration may be advantageous
for studying strongly nonlinear regimes well above the dynamo threshold.

Another important difference concerns the direction of the axial flow
relative to the agitator. In the Riga dynamo, a propeller pushes the
liquid sodium axially. In contrast, in our setup, the impeller produces
a centrifugal radial jet while simultaneously drawing axial flow toward
itself. At the opposite end of the cylinder, a flow topology forms
that resembles a tornado or, to a lesser extent, an accretion disk
with polar jets. These distinctions, however, do not alter the fundamental
generation mechanism of the magnetic field by the helical vortex core.

The flow in the von K\'{a}rm\'{a}n sodium (VKS) dynamo experiment\citep{Monchaux2009}
is driven by two relatively large-diameter impellers. The liquid-metal
vessel has a diameter nearly equal to its height, and the impellers
counter-rotate. The design philosophy of the VKS experiment emphasizes
strong turbulence rather than screw-like coherent motion. As a result,
both the flow structure and the overall dynamo concept differ substantially
from the Ponomarenko-type dynamo considered here.


The paper is organized as follows. In the next section, we introduce the mathematical model and briefly describe the numerical method, which is based on the Chebyshev-tau approximation. Results of the direct numerical simulations of the impeller-driven flow are presented in Sec. \ref{subsec:DNS}, and these velocity fields are then used in Sec. \ref{subsec:Dyn} to determine the threshold of dynamo action by numerically integrating the induction equation. The paper concludes with a discussion of the results in Sec.~\ref{sec:Disc}.

\section{Model}

We consider  a swirling flow of an incompressible fluid with electrical
conductivity $\sigma$, which is driven by an impeller in a finite-length
cylinder of radius $R_{0}$ and focus on the middle part of the cylinder,
where the axial variation of the flow is relatively small. The cylinder
is surrounded by an electrically insulating medium, and the impeller
is modeled by an axially symmetric azimuthal body force distribution
representing the time-averaged action of a rotating permanent magnet
placed coaxially near one end of the cylinder.

Following the classical kinematic dynamo approach, the magnetic field
is supposed to be generated by the mean flow, with turbulent fluctuations
playing no significant role in this process. The flow is represented
in cylindrical coordinates $(r,\phi,z)$ by an axially and rotationally
invariant velocity profile $\vec{v}(r)=\vec{e_{\phi}}r\Omega(r)+\vec{e}_{z}W(r),$
which is obtained by averaging the time-dependent 3D numerical solution
of the Navier-Stokes equation. 

Using $R_{0}$ and $\mu_{0}\sigma R_{0}^{2}$ as the length and time
scales, the induction equation for the magnetic field $\vec{B}(\vec{r},t)$
in the cylinder can be written in the following dimensionless form
\global\long\def\Rm{\textit{Rm}}%
\begin{equation}
\partial_{t}\vec{B}=\Rm\vec{\nabla}\times\left(\vec{v}\times\vec{B}\right)+\vec{\nabla}^{2}\vec{B},\label{eq:ind}
\end{equation}
where $\Rm=\mu_{0}\sigma R_{0}V_{0}$ is the magnetic Reynolds number
based on the maximal axial velocity $V_{0},$ which corresponds to
$\max_{r}|W(r)|=1$ in the dimensionless form. As $\vec{v}$ depends
only on $r$, particular solutions of the magnetic field can be sought
as the normal mode  
\global\long\def\I{\mathrm{i}}%
\begin{equation}
\vec{B}(r,\phi,z,t)=\vec{\hat{B}}(r,t)e^{\I(kz+m\phi)}+\text{c.c.},\label{Bprtb}
\end{equation}
with a complex time-dependent radial amplitude distribution $\vec{\hat{B}}(r,t)$.
The magnetic field is completely determined
by any two of its components due to solenoidality. In our case, it is advantageous to employ
the radial and azimuthal components as the independent variables.
Combining the respective components of Eq. (\ref{eq:ind}), it can be written in the following compact form:
\begin{equation}
\partial_{t}\hat{B}_{\pm}=\left[D_{m\pm1}-k^{2}-\I\Rm(m\Omega+kW)\right]\hat{B}_{\pm}\pm\I\Rm r\Omega'(\hat{B}_{+}+\hat{B}_{-})/2,\label{bpmu}
\end{equation}
where $\hat{B}_{\pm}=\hat{B}_{r}\pm i\hat{B}_{\phi}$ and $D_{m}=\frac{d^{2}\,}{dr^{2}}+\frac{1}{r}\frac{d\,}{dr}-m^{2}r^{-2}.$
Outside the cylinder $(r>1)$, where $\sigma=0$, the induction equation
reduces to $D_{m\pm1}\hat{B}_{\pm}=0$ and has an analytical solution:
\[
\hat{B_{\pm}}(r)=A_{\pm}K_{m\pm1}(kr),
\]
where $K_{m}(x)$ is the modified Bessel function of the second kind
and $A_{\pm}$ are two  unknown constants. The irrotationality of the
free-space magnetic field results in $A_{+}=A_{-}.$ Using this analytic
solution and the continuity of the magnetic field, we obtain the two
boundary conditions for $\hat{B}_{\pm}$ 
expressed as 
\begin{align}
\hat{B}_{+}-\gamma\hat{B}_{-} & =0,\label{BC-gamma}\\
\hat{B}'_{+}+\hat{B}'_{-}+s\hat{B}_{-} & =0,\label{BC-s}
\end{align}
where $\gamma=K_{m+1}(k)/K_{m-1}(k)$ and $s=\gamma(m+1)-(m-1)+(\gamma-1)k^{2}/m$.
The last condition results from the continuity of the axial component
of the magnetic field and ensures that no electric current flows through
the cylinder boundary at $r=1.$ At the axis $(r=0),$ for $m=1$
considered in the following, we have the pole condition

\begin{equation}
\hat{B}_{+}=0,\label{bc:bp0}
\end{equation}
which follows from the single-valuedness of the magnetic field. The formulation of the problem
above is equivalent to the one used by Dobler \textit{et al.} \citep{Dobler2003}

Equations (\ref{bpmu}) are integrated over time numerically using the Adams-Bashforth
method, which results in a system of two ordinary differential equations
\begin{equation}
D_{m\pm1}\hat{B}_{\pm}^{n+1}-(k^{2}+1.5/\Delta t)\hat{B}_{\pm}^{n+1}=2F_{\pm}^{n}-F_{\pm}^{n-1}-(2\hat{B}_{\pm}^{n}-0.5\hat{B}_{\pm}^{n-1})/\Delta t,\label{bpm}
\end{equation}
where $\hat{B}_{\pm}^{n}(r)=\hat{B}_{\pm}(r,t_{n})$, $\Delta t=t_{n+1}-t_{n}$
, and $F_{\pm}$ represent the convective terms with $\Rm$ in Eq.
(\ref{bpmu}). The problem is solved numerically using the Chebyshev
tau approximation, where for $m=1$ the solution is sought as a series
of even Chebyshev polynomials  
\begin{equation}
\hat{B}_{\pm}^{n}(r,t_n)=\sum_{i=0}^{N}b_{i}^{\pm}(t_{n})T_{2i}(r),\label{Cheb-bpm}
\end{equation}
with unknown time-dependent coefficients $b_{i}^{\pm}(t_{n})$. This
series contains only even Chebyshev polynomials because of the axial
reflection symmetry, which implies $\hat{B}_{\pm}$ for mode $m=1$
to be even functions of $r.$

Substituting (\ref{Cheb-bpm}) into (\ref{bpm}) and using the orthogonality
of Chebyshev polynomials, we obtain two sets of $N+1$ linear algebraic
equations for $b_{i}^{\pm}.$  The equations are modified by the boundary
conditions (\ref{BC-gamma},\ref{BC-s}), which take the form: 
\begin{align}
\sum_{i=0}^{N}(b_{i}^{+}+\gamma b_{i}^{-}) & =0,\\
\sum_{i=0}^{N}\left(4i^{2}(b_{i}^{+}+b_{i}^{-})+sb_{i}^{-}\right) & =0.
\end{align}
According to the tau method, these boundary conditions override the
last two algebraic equations for $b_{i}^{\pm}$, which are produced
by the projection of (\ref{bpm}) onto $T_{2N}(r).$ Likewise, the second
to last equation for $b_{i}^{+}$ is overridden by the pole condition
(\ref{bc:bp0}), which takes the form: 
\begin{equation}
\left.\hat{B}_{+}\right|_{r=0}=\sum_{i=0}^{N}(-1)^{i}b_{i}^{+}=0.
\end{equation}
Note that there is another pole condition $\hat{B}'_{-}=0$ at $r=0$,
which is satisfied by (\ref{Cheb-bpm}) automatically.

Computations are carried out using the spatial resolution $N=96$
and the time step $\Delta t=0.01\Rm^{-1}$, starting from the initial
state $\hat{B}_{+}=0$ and $\hat{B}_{-}$, defined by three non-zero
coefficients $b_{0,1,2}^{-}=(3,-4,1)/6$, which satisfy the boundary conditions
(\ref{BC-gamma},\ref{BC-s}) at $t=0$, as both $\hat{B}_{-}$ and its radial
derivative are zero at $r=1$. 
The leading eigenvalue
is computed using the Goldhirsch method\citep{Goldhirsch1987} based
on five trailing snapshots, each separated by 20 time steps. The calculation
proceeds until the change in the leading eigenvalue between successive
iterations falls below 0.01\%. Accuracy to three decimal places is
verified by repeating the computation with twice the temporal resolution
and with $N=128$.

The code was validated by solving a test problem for $k=2$ and
\[
\Omega(r)=-2W(r)=\Rm(1-\tanh(\alpha(r-1/2))).
\]
The critical value $\Rm=41.3$ agrees almost perfectly
with the analytical result $\Rm=41.32$ for the classical Ponomarenko
dynamo, which corresponds to the limit $\alpha\to\infty$. The oscillation
frequency at onset was $17.55$ and $17.85$ for $\alpha=20$ and
$50$, respectively, corresponding to deviations of $2.2\%$ and $0.5\%$
from the analytical value $17.94.$

The velocity profiles $\Omega(r)$ and $W(r)$ are obtained by direct
numerical simulation (DNS) using the spectral code described by
\citet{Grants2021}. This code solves the three-dimensional, transient,
incompressible Navier-Stokes equations in a cylindrical domain with a 
prescribed body force. The force is generated either by a coaxial,
radially magnetized rotating cylindrical permanent magnet or by a rotating
magnetic dipole. Its spatial distribution is computed by neglecting both
the skin effect and the influence of the flow.\citep{Berenis2020}

\begin{figure}
\centering
\includegraphics[width=0.5\textwidth]{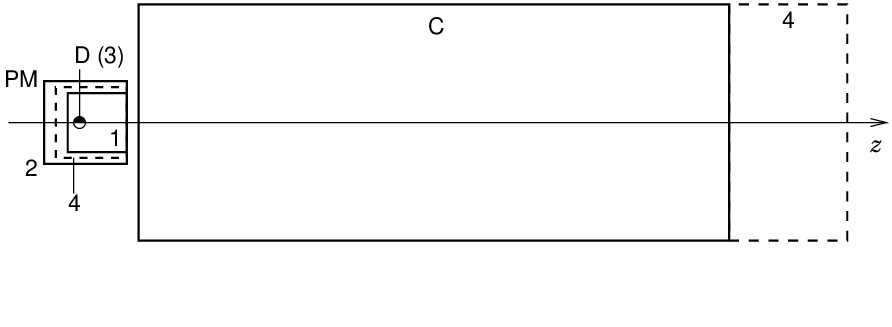}\put(-195,25){(a)} 
\includegraphics[width=0.5\textwidth]{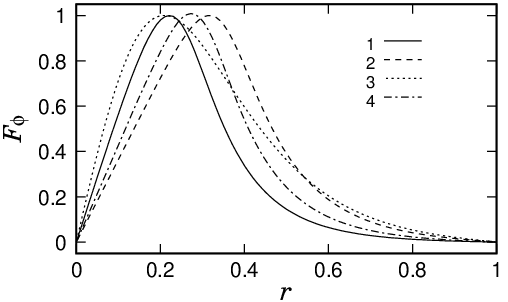}\put(-195,25){(b)}
\caption{\label{scheme:fig} Schematic of the four numerically simulated configurations
(a) and the corresponding radial profiles of the azimuthal body force
at the magnet-facing surface (b). The flow in the liquid cylinder
(C) is driven either by a rotating permanent magnet (PM) or by a rotating
magnetic dipole (D). The cylinder has a length-to-diameter ratio of
2.5 in cases 1--3 and 3.0 in case 4 (shown with dashed lines). The
permanent magnet has a diameter and length equal to 0.25 and 0.35
cylinder diameters in cases 1 and 2, respectively, and 0.3 cylinder
diameters in case 4. The gap between the PM and the liquid is 0.05
cylinder diameters. In case 3, the dipole is positioned 0.25 cylinder
diameters away from the liquid.}
\end{figure}

The core flow described by $\Omega(r)$ and $W(r)$ is weakly affected 
by the distribution of the driving body force because the two are
spatially separated. Specifically, the body force is negligible in
the central part of the cylinder, where the flow is primarily shaped 
by vortex stretching resulting from the centrifugal action of the
impeller. The magnet arrangement in the four configurations is shown
schematically in Fig. \ref{scheme:fig}(a). The chosen aspect ratio
of the cylindrical container is close to that of commercial sodium storage
tanks.

Time-averaged distributions of driving force are computed using the analytical
expressions from \citet{Berenis2020} and are displayed in Fig.
\ref{scheme:fig}(b), with maxima scaled to unity. The flow is time-averaged
over approximately $50$ vortex core revolutions, following
an initial transient of comparable duration.

Simulations are performed with a spatial resolution of $97 \times 97 \times 109$ modes. To evaluate the influence of numerical parameters, velocity profiles are also computed using a coarser resolution of $83 \times 83 \times 97$ modes. The temporal resolution corresponds to approximately 5000 time steps per vortex-core revolution.

\section{Results}

\subsection{\label{subsec:DNS} Velocity field }

Time-averaged velocity distributions were computed numerically using
a constant azimuthal body force with the radial profiles shown in
Fig. \ref{scheme:fig}(b). The force distribution was multiplied by a 
dimensionless factor, chosen to be as large as possible while still 
yielding numerically reliable velocity distributions. The numerical error was estimated
by comparing the $z$-averaged velocity profiles obtained at two spatial
resolutions ($97\times97\times109$ and $83\times83\times97$ modes).
The difference between these profiles was less than $5\%$ of the
corresponding maximum value when the flow Reynolds number, based on 
the maximum absolute velocity and the cylinder radius, was
2110, 1660, 1400, and 2120 in cases 1--4, respectively.

\begin{figure}
\centering
(a) \includegraphics[height=0.6\textheight]{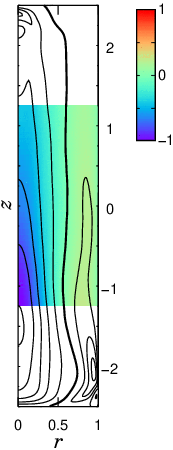}\qquad 
(b) \includegraphics[height=0.6\textheight]{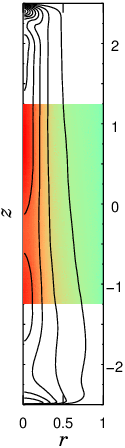}
\caption{ \label{V2d:fig} Axial (a) and angular (b) velocity distributions for case 1. Colors are superimposed in the central half of the cylinder, over which the profiles in Fig. \ref{Vp:fig} are averaged. The isoline interval is 0.2 times the corresponding maximum absolute value in the averaging domain.}
\end{figure}

Figure \ref{V2d:fig} shows a representative distribution of angular and axial velocity. 
In all four cases, the distributions showed a similar degree of axial non-uniformity in the middle half of the cylinder, the region over which the velocity profiles were averaged. Axial variations were most pronounced in case 4, which involved a longer cylinder with a diameter-to-height ratio of $0.33$. The mean radial velocity in the central region of the cylinder was negligible in all cases.

\begin{figure}
\centering
\includegraphics[width=0.5\textwidth]{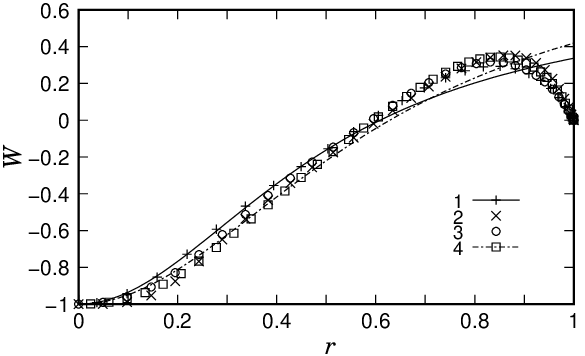}\put(-25,25){(a)} 
\includegraphics[width=0.5\textwidth]{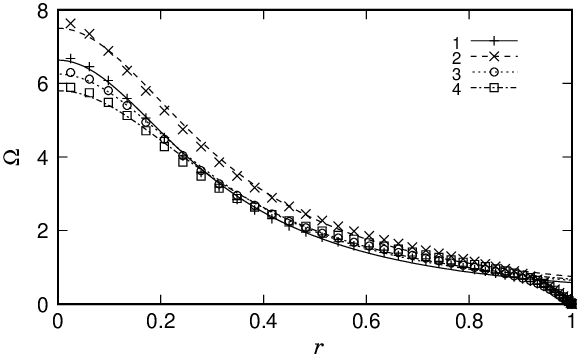}\put(-205,25){(b)}

\caption{ \label{Vp:fig} Axial (a) and angular (b) velocity profiles averaged
over $z$ in the central half of the cylinder. The numbers in the legend correspond 
to the cases shown in Fig. \ref{scheme:fig}.
The velocity profile fits defined by Eq. (\ref{Vprofiles}) are shown as curves, and the corresponding fit parameters are listed in Table \ref{fit:table}. }
\end{figure}

Figure \ref{Vp:fig} shows the velocity profiles averaged over $z$
in the middle half of the cylinder and scaled by the maximum absolute
axial velocity. The axial velocity profiles are remarkably similar
despite the significantly different forcing distributions and the
variations in the Reynolds number. 
Outside the boundary layer at the side wall ($r=1$), all four numerically
computed profiles are well approximated by the following expressions

\begin{align}
\Omega(r) & =a b\, g(a r),\label{Vprofiles}\\
W(r) & =\frac{g(c r)-d}{d-1},\label{Wprof}
\end{align}
where $g(x)=(1+x^{2})^{-1}$ and $a$, $b$, $c$ are free parameters
with $d=\ln(1+c^{2})/c^{2}$ ensuring a zero net flow rate $\int_{0}^{1}W(r)r\thinspace dr=0.$

\begin{table}[b]
\centering
\caption{ \label{fit:table} Fitted parameters of the velocity profiles for the four numerically simulated cases.}
\begin{tabular}{llll}
\hline 
 & $a$  & $b$  & $c$ \tabularnewline
\hline 
1  & 3.22  & 2.06  & 2.01 \tabularnewline
2  & 3.00  & 2.50  & 1.65 \tabularnewline
3  & 2.95  & 2.12  & 1.82 \tabularnewline
4  & 2.75  & 2.11  & 1.63 \tabularnewline
\hline 
\end{tabular}
\end{table}

\subsection{\label{subsec:Dyn} Dynamo}

\begin{figure}
\centering
\includegraphics[width=0.5\textwidth]{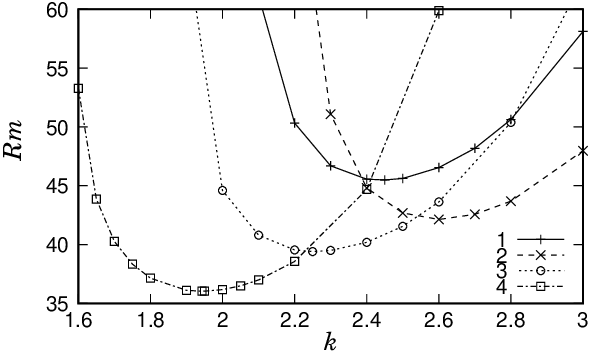}\put(-200,25){(a)} 
\includegraphics[width=0.5\textwidth]{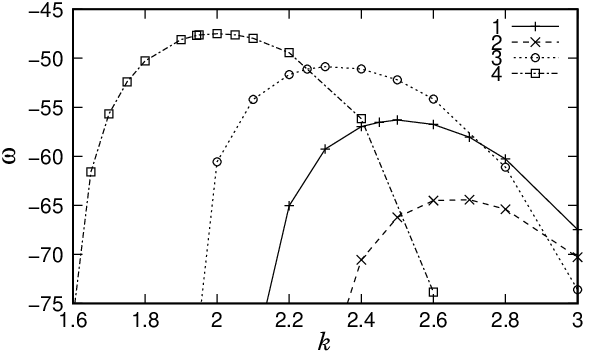}\put(-200,25){(b)} 
\caption{\label{Rmc_DNS:fig} 
Magnetic Reynolds number (a) and oscillation frequency (b) of marginal modes of the 
magnetic field versus axial wavenumber $k$ and $m=1$ for the four velocity profiles corresponding to the configurations shown in Fig. \ref{scheme:fig}.}
\end{figure}

In this section, we numerically simulate the temporal evolution of the
magnetic field in the four flow configurations whose velocity profiles are
shown in Fig. \ref{Vp:fig}. We use the `snapshot' method \citep{Goldhirsch1987}
to determine the leading eigenvalue $\lambda$ for the normal mode
of the magnetic field (\ref{Bprtb}) with $m=1$ by integrating over
time the induction equation (\ref{bpmu}) with computed velocity profiles
for various magnetic Reynolds numbers $\Rm$ and axial wavenumbers
$k$. The marginal $\Rm$, by exceeding which the magnetic field mode
with the axial wavenumber $k$ starts to grow exponentially at
a rate $\Re(\lambda)>0$, is plotted in Fig. \ref{Rmc_DNS:fig}(a).
The imaginary part of the leading eigenvalue, $\Im(\lambda)=\omega$, 
defining the frequency with which the magnetic field oscillates in
time at this point, is shown in Fig. \ref{Rmc_DNS:fig}(b). The
lowest point on each marginal $\Rm$ curve defines the critical magnetic
Reynolds number, $\Rm_{c}$, at which the given flow becomes able to sustain
a magnetic field with the corresponding wavenumber and frequency. Critical
parameters computed for the velocity profiles obtained by DNS are
summarized in Table \ref{DNS-vp:table} along with the results for
the corresponding best-fit profiles Eq.(\ref{Vprofiles}). For
the fitted profiles, the critical magnetic Reynolds numbers $\Rm_{c}$
are about 15--50\% higher than those for the DNS counterparts.
The fitted velocity profiles also produce slightly larger critical
wavenumbers $k_{c}$ and frequencies $\omega_{c}$. Small variations
in the velocity profiles (Fig. \ref{Vp:fig}) are seen to cause much
bigger variations in $\Rm_{c}$, demonstrating the strong sensitivity
of the dynamo threshold to the velocity field.

Note that the marginal modes of the magnetic field, which by definition
have a zero temporal growth rate $\Re(\lambda)=0$, generally have 
a non-zero frequency $\Im(\lambda)=\omega.$ As seen in Fig. \ref{Rmc_DNS:fig},
this frequency is not constant but varies with the wavenumber $k.$
As a result, the modes have not only a non-zero phase velocity $\omega/k,$
but also a non-zero group velocity $\partial_{k}\omega.$ Note that the
latter is the velocity of the wave packet made up of a superposition of modes
centered around $k.$ Such a growing wave packet of the modes centered
around $k_{c}$ emerges as soon as $\Rm$ exceeds $\Rm_{c}.$ When
this wave packet with the growing magnetic field travels along the
cylinder with a non-zero group velocity, the magnetic field grows
only in the co-moving frame of reference, eventually decaying at any fixed
position. For the magnetic field to persist in the laboratory (fixed)
frame of reference, the group velocity of the critical mode must be
zero. This means that, at the critical wavenumber $k_{c}$, where marginal
$\Rm$ attains its minimum and thus $\partial_{k}\Rm=0$, we also need
$\partial_{k}\omega=0$, which implies a local extreme of $\omega$
at the same point. 

Numerical results show that $\omega(k)$ has a maximum that is slightly offset from that of $\Rm(k).$ This means that the critical mode has a non-zero group velocity $v_{g}=\partial_{k}\omega/\Rm$. Adding this $v_{g}$ to the axial velocity profile $W(r)+v_g $ shifted the maximum of $\omega$ to $k_{c}.$ 

\begin{figure}
\centering
\includegraphics[width=0.5\textwidth]{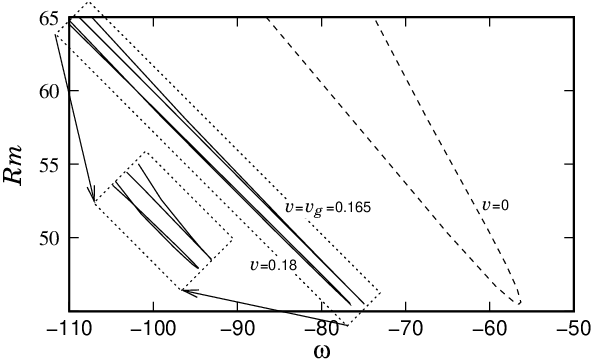}
\caption{ \label{Rm-w:fig} The marginal $Rm$ as a function of marginal frequency $\omega$ for case 1 with a uniform back-flow $v$. }
\end{figure}

When the marginal $\Rm$ is represented parametrically as a function
of the marginal frequency $\omega(k):$ 
$\Rm = \Rm(\omega(k)),$
at the point of absolute instability, where $\partial_k\Rm = \partial_k\omega = 0$ and  $\omega_{c}=\omega(k_{c}),$ we have
\[
\partial_{\omega}\Rm=\partial_{k}\Rm/\partial_{k}\omega=
\partial_{k}^{2}\Rm/\partial_{k}^{2}\omega.
\]
As this point is a minimum of $\Rm,$ in general, we have $\partial_{k}^{2}\Rm>0$
and, thus, according to the relation above $\partial_{\omega}\Rm\not=0.$ This non-zero derivative, in turn, means that the marginal $\Rm(\omega(k))$ curve at the minimum has a cusp -- a singularity in the form of a sharp peak with coincident tangents.\citep{Priede2009}
This is a characteristic feature of absolute instability, which
was originally identified in the complex frequency plane.\citep{Kupfer1987}
When a uniform back-flow exceeding the group velocity is added to
the axial velocity profile, the cusp morphs into a closed loop with
a self-intersection point on the marginal $\Rm(\omega(k))$ curve.
This self-intersection point is a prerequisite of absolute instability
at complex wave numbers.\citep{Priede1997} Our current results
show such an intersection point (Fig.\ \ref{Rm-w:fig}), and thus absolute instability appears
only at sufficiently strong additional back-flow.

\begin{figure}
\centering
\includegraphics[width=0.5\textwidth]{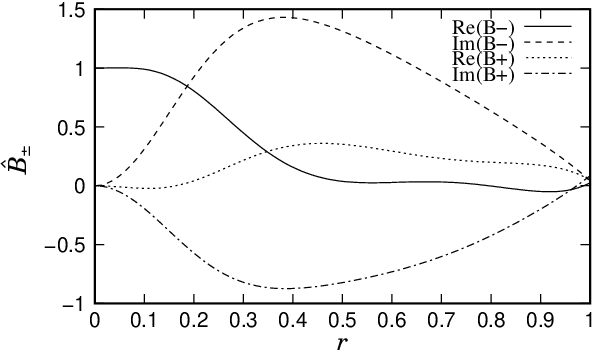}
\caption{ \label{b1:fig} The critical mode of the magnetic field for case 1 normalized so that $\hat{B}_-(0)=1$. }
\end{figure}

Figure \ref{b1:fig} shows the radial profile of the critical magnetic-field mode for case 1. The field reaches its maximum amplitude at $r \approx 0.4$ and decreases to nearly zero at the cylinder boundary, $r = 1$. A similar radial distribution is found for all other cases.
The low magnitude of the critical mode at the boundary $r=1$ makes it an almost invisible dynamo. The invisible  dynamo produces no magnetic field outside the conducting domain. \citep{Simkanin2008}  This study\citep{Simkanin2008} sought an invisible dynamo in an infinite cylinder and only decaying invisible modes have been observed.
Instead of requiring strictly invisible dynamo one could look for a velocity field that generates magnetic field with a possibly low
magnitude at the boundary (relative to interior).   Our results suggest that such weak formulation might produce practically
meaningful results.

\begin{table}
\centering{}
\caption{\label{DNS-vp:table} Threshold parameters in the four numerically
simulated cases. The asterisk denotes the results obtained using the corresponding best-fit velocity profiles (\ref{Vprofiles}). }
\begin{tabular}{llllllll}
\hline 
 & $\Rm_{c}$ & $k_{c}$  & $\omega_{c}$ & $v_{g}$  & $\Rm_{c}^{*}$ & $k_{c}^{*}$ & $\omega_{c}^{*}$ \tabularnewline
\hline 
1  & 45.5  & 2.45  & -56.5  & 0.165  & 52.9  & 2.84  & -59.9 \tabularnewline
2  & 42.1  & 2.60  & -64.5  & 0.188  & 64.0  & 3.02  & -91.4 \tabularnewline
3  & 39.4  & 2.25  & -51.1  & 0.182  & 47.7  & 2.59  & -57.3 \tabularnewline
4  & 36.0  & 1.95  & -47.6  & 0.190  & 44.1  & 2.33  & -53.6 \tabularnewline
\hline 
\hline 
 &  &  &  &  &  &  & \tabularnewline
\end{tabular}
\end{table}

\section{\label{sec:Disc}Discussion}

This study shows that an impeller-driven liquid sodium flow in a
cylinder can reach the convective instability threshold for the generation
of a magnetic field. For instance, the critical magnetic Reynolds number $\Rm_{c}=36$
found for the fourth considered flow configuration corresponds to
$V_{0}\approx5.4$ m/s in a 4 m$^{3}$ liquid sodium tank with a diameter
$2R_{0}\approx1.2$ m and 3:1 length-to-diameter ratio. This is
somewhat lower than the velocity in the Riga dynamo
of similar length, which used about 1.5 m$^3$ of
sodium. In a 22 m$^{3}$ tank, we expect the critical velocity to be below 3 m/s, which should allow the setup to reach a strongly supercritical dynamo regime. 

However, these results are not fully conclusive because they are based on a highly idealized model of an infinitely long cylinder. In this model, the growing magnetic field propagates along the cylinder with a non-zero group velocity, which is characteristic of a convective instability. This means that although the flow can amplify the magnetic field, it cannot sustain it indefinitely because the field eventually escapes through the end of the cylinder.

Several approaches may help to overcome this limitation. First, only
a few representative configurations have been explored here. At least
two geometric parameters can be tuned: the cylinder aspect ratio and
the impeller-to-cylinder diameter ratio. Additional parameters arise
when the flow is driven by a mechanical impeller. These degrees
of freedom may allow for a configuration in which the
group velocity vanishes. The velocity profile approximation (\ref{Vprofiles})
is well suited for such a search. In this parametrization, $a$ and
$c$ characterize the widths of the angular and axial core velocity
profiles, respectively, while $b$ defines the pitch of the helical
core flow.

The convective instability can, in principle, be made self-sustained
by magnetically coupling the two ends of the cylinder so that the
growing magnetic field arriving at one end is fed back to the other end.
Such a mechanism could be realized, for example, by means of an external
coil system. A conceptually similar approach was proposed by \citet{Bourgoin2006}, where an additional amplifier was incorporated into an electrical feedback loop.

Another possibility is to employ a second, identical cylinder placed
side by side. If the axial flows in the two cylinders are directed
oppositely, the combined system would possess no preferred axial direction.
Magnetic coupling through the ends of the cylinders could then provide
the necessary feedback without significantly increasing the critical
magnetic Reynolds number $\Rm$.

A further potential mechanism relies on the radially converging boundary
layer, which may supply the feedback required to sustain amplification
by the core helical flow. Such coupling may be facilitated by the
relatively short axial wavelength of the marginal magnetic mode: in
all four simulated cases, approximately two full critical wavelengths
fit within the cylinder.
If effective, such feedback could be
captured by two-dimensional dynamo model in a finite cylinder \citep{Stefani1999,Xu2008} with vacuum magnetic boundary conditions. Two-dimensional modeling would also clarify influence of the magnetic boundary conditions at the ends of a finite-length cylinder.

The convective nature of the dynamo is not the only potential obstacle
to realizing the considered configuration as a working laboratory
set-up. Our results indicate that the amplification of the magnetic
field is sensitive to relatively small variations in the velocity
profile. The present simulations were performed at hydrodynamic Reynolds
numbers of 1000--2000, which effectively corresponds to an unrealistically 
large magnetic Prandtl number of order $0.02$. In reality, the velocity 
distributions may differ substantially from those obtained here.

Nevertheless, the observed $\propto r^{-2}$ dependence should remain
robust, at least for the angular velocity profile, because
it follows directly from angular momentum conservation in a stretched
vortex. The axial velocity profile is influenced in part by Ekman
pumping in the radially converging boundary layer. For this reason,
an approximate $\propto r^{-2}$ scaling may also hold in the axial
velocity component at much higher Reynolds numbers. Moreover, the
axial uniformity of the bulk flow is expected to improve with increasing
Reynolds number due to the Taylor--Proudman theorem. Thus, the core
velocity field is likely to retain the general form (\ref{Vprofiles}),
parameterized by three characteristic scales.

Among these, the pitch of the helical core flow, $b^{-1}$, appears
particularly important. Our DNS results yielded pitch values around $0.5$,
which are significantly smaller than the optimal value $\approx1.3$ for the
classical Ponomarenko dynamo \citep{Gailitis1976}. The pitch is likely
to be even lower under realistic turbulent sodium flow conditions,
in which case the critical magnetic Reynolds number $\Rm_{c}$ could
become prohibitively large for laboratory realization.
The observed sensitivity of $\Rm_c$ to velocity profile shape requires accurate velocity measurements 
at relevant hydrodynamical conditions as an important first step towards the laboratory dynamo.
If a too small core pitch value is measured, then the magnetic impeller could be replaced by
%
a mechanical impeller specifically designed to enhance axial flow. In practice, this could be implemented with a propeller operated in reverse, driving axial flow toward the adjacent end wall. 

\section*{Author declarations}

\subsection*{Conflict of Interest}
The authors have no conflicts to disclose.

\subsection*{ Data availability }
The data that support the findings of this study are available from the corresponding author upon reasonable request.

\bibliography{dyn/dyn}

\end{document}